\documentclass[fleqn,10pt]{wlscirep}

\title{Non-blinking single-photon emitters in silica}

\author[1]{Freddy T. Rabouw}
\author[2]{Nicole M. B. Cogan}
\author[1]{Anne C. Berends}
\author[1]{Ward van der Stam}
\author[1]{Daniel Vanmaekelbergh}
\author[3]{A. Femius Koenderink}
\author[2]{Todd D. Krauss}
\author[1,*]{Celso de Mello Doneg\'{a}}
\affil[1]{Debye Institute for Nanomaterials Science, Utrecht University, Princetonplein 5, 3584 CC Utrecht, The Netherlands}
\affil[2]{Department of Chemistry, University of Rochester, Rochester, New York 14627, United States of America}
\affil[3]{Center for Nanophotonics, FOM Institute AMOLF, Science Park 104, 1098 XG Amsterdam, The Netherlands}

\affil[*]{c.demello-donega@uu.nl}

\begin{abstract}
Samples for single-emitter spectroscopy are usually prepared by spin-coating a dilute solution of emitters on a microscope cover slip of silicate based glass (such as quartz). Here, we show that both borosilicate glass and quartz contain intrinsic defect colour centres that fluoresce when excited at 532 nm. In a microscope image the defect emission is indistinguishable from spin-coated emitters. The emission spectrum is characterised by multiple peaks, most likely due to coupling to a silica vibration with an energy of 160--180 meV. The defects are single-photon emitters, do not blink, and have photoluminescence lifetimes of a few nanoseconds. Photoluminescence from such defects may previously have been misinterpreted as originating from single nanocrystal quantum dots.
\end{abstract}

\begin{document}

\flushbottom
\maketitle

\thispagestyle{empty}

\section*{Introduction}

Single-emitter spectroscopy is essential to build a comprehensive microscopic picture of the physical processes involved in fluorescent emitters. In experiments on ensembles of (slightly) inhomogeneous emitters, many processes are hidden by averaging. For example, only single-emitter experiments have been able to reveal that the emission from many types of fluorescent species exhibits spectral diffusion \cite{Ambrose1991}, blinking \cite{Nirmal1996,Vandenbout1997}, and anti-bunching \cite{Fleury2000,Lounis2000}. \\

The photophysical properties of single molecule emitters such as organic dye molecules \cite{Ambrose1991,Vandenbout1997,Lounis2000}, and colloidal quantum dots \cite{Nirmal1996,Mason1998,English2002,Shimizu2002,Martin2008,Chizhik2009,Spinicelli2009,Wang2009,Kusova2010,Galland2012,Schmidt2012,Rabouw2013} have been examined extensively for the past two decades. Samples for single molecule photoluminescence spectroscopy are usually prepared by spin-coating a dilute solution of the emitters from a liquid or polymer solution onto a glass or quartz cover slip \cite{Nirmal1996,Mason1998,English2002,Martin2008,Chizhik2009,Wang2009,Spinicelli2009,Schmidt2012,Rabouw2013}. The surface density of emitters must be sufficiently low ($<$1 $\mu$m$^{-2}$), so that the diffraction-limited excitation spot of a continuous-wave (cw) or pulsed laser can address an individual molecule. Most studies have focused on relatively bright emitters, with large absorption cross-sections, high quantum efficiencies and/or short photoluminescence lifetimes. Experiments on dimmer emitters (such as Si \cite{Mason1998,English2002,Martin2008,Chizhik2009,Kusova2010,Schmidt2012}, InP \cite{Dennis2012}, InAs \cite{Bischof2014}, or CuInS$_2$ (see below)) are more challenging, because the signal is easily obscured by background counts from fluorescence of the substrate, laser reflections, and detector dark counts. For a proper interpretation of experimental data on relatively dim emitters, it is important to understand the origin of background signals. \\

Here we show that microscope cover slips of silica (glass or quartz), commonly used for single-emitter studies, contain intrinsic single-photon emitting centres. Photoluminescence (PL) from such centres is observed under excitation at 532 nm, both from cover slips of borosilicate glass and fused quartz. The centres have a characteristic emission spectrum with two or three Gaussian peaks, and PL lifetimes of a few nanoseconds. Single-photon emitters with very similar characteristics have been observed before in several studies \cite{Mason1998,English2002,Martin2008,Chizhik2009,Wang2009,Kusova2010} where silica cover slips were used as a substrate to spin-coat the sample. While in these studies the PL was ascribed to nanocrystals (of various types), our results imply that it may in fact have originated from luminescent centres intrinsic to the silica substrate.

\section*{Results}

We investigate the fluorescence from bare silica microscope cover slips (borosilicate glass or fused quartz) cleaned with hydrochloric acid, followed by isopropanol. 
The cover slips were excited with a pulsed (10 MHz) Nd:YVO$_4$ laser (532 nm). See Methods for details. \\

\begin{figure}[ht]
\centering
\includegraphics[width=\linewidth]{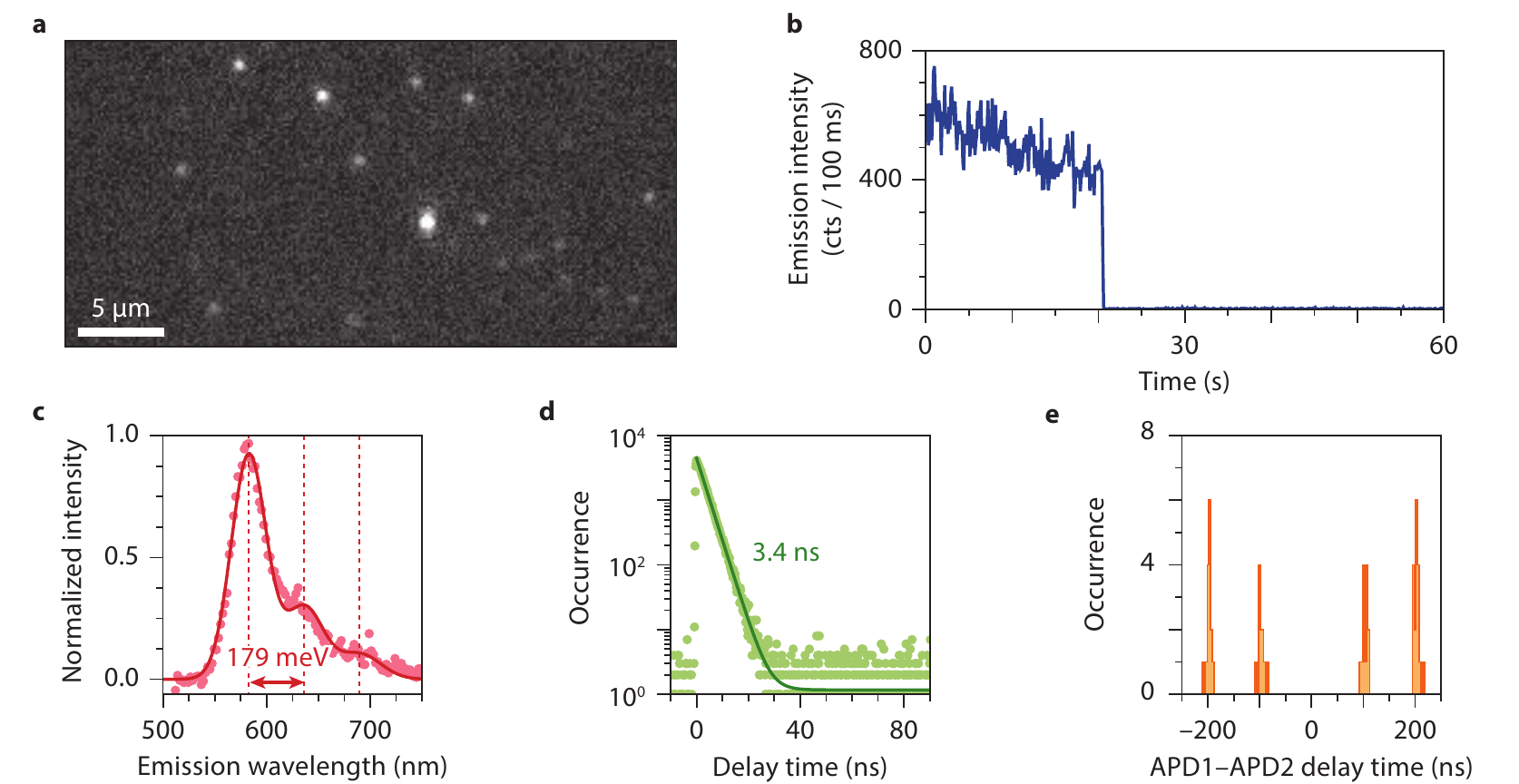}
\caption{
\textbf{Single-emitter luminescent centres on a microscope glass cover slip.}
\textbf{(a)} A fluorescence image of a bare borosilicate glass cover slip epi-illuminated with the laser (532 nm; 200 $\mu$W) defocussed to a $\sim$50 $\mu$m diameter spot reveals luminescent spots at a surface density of 0.01--0.1 $\mu$m$^{-2}$.
\textbf{(b)} A typical intensity time trace of a single spot. The emission is non-blinking, but the spot bleaches after 21 s.
\textbf{(c)} The emission spectrum of the spot shows a main peak in the orange with side bands at longer wavelengths. The solid line is a fit to a series of three Gaussians regularly spaced at a separation of 179 meV.
\textbf{(d)} The PL decay trace of the spot is single exponential, with a fitted PL lifetime of 3.4 ns (solid line).
\textbf{(e)} The two-photon correlation function demonstrates anti-bunching, indicating that we are observing a single-photon emitter.
\label{fig:borosilicate}
}
\end{figure}

Fig.~\ref{fig:borosilicate} gives an overview of the properties of the luminescent defects we found in microscope cover slips of borosilicate glass. Fig.~\ref{fig:borosilicate}a is an epi-fluorescence wide-field microscopy image of a bare glass cover slip. Isolated luminescent spots are clearly visible with a density of 0.01--0.1 $\mu$m$^{-2}$. The image closely resembles a typical fluorescence image of single molecules or single quantum dots deposited on a substrate. The substrate observed in Fig.~\ref{fig:borosilicate} is however bare. We conclude that the PL must originate from luminescent centres in or on the substrate itself. Focusing the laser excitation to a diffraction limited spot, we can investigate individual centres. Fig.~\ref{fig:borosilicate}b presents a typical intensity trace of a centre, showing that the emission is non-blinking. We find that most centres photobleach within a minute (as the one in Fig.~\ref{fig:borosilicate}b), while only a few remain emissive for several minutes. \\

The emission spectrum of the centre is presented in Fig.~\ref{fig:borosilicate}c. The PL peaks at 583 nm, but is clearly asymmetric with strong sidebands to the red of the main peak. The PL spectrum can be fitted to a progression of three Gaussians at regular energy separation (solid line), yielding a peak separation of 179 meV. The time-resolved PL decay curve of the centre (Fig.~\ref{fig:borosilicate}d) is single-exponential over three orders of magnitude in dynamic range, with a PL lifetime of 3.4 ns (solid line). In Fig.~\ref{fig:borosilicate}e we plot the two-photon correlation function $g^{(2)}$, i.e. the probability distribution of delay times between consecutive photon detection events \cite{Nair2011,Cui2014}. The absence of coincidence counts at zero delay proves that the emission is anti-bunched, i.e. the luminescent centre emits no more than a single photon per excitation pulse. \\

\begin{figure}[ht]
\centering
\includegraphics[width=\linewidth]{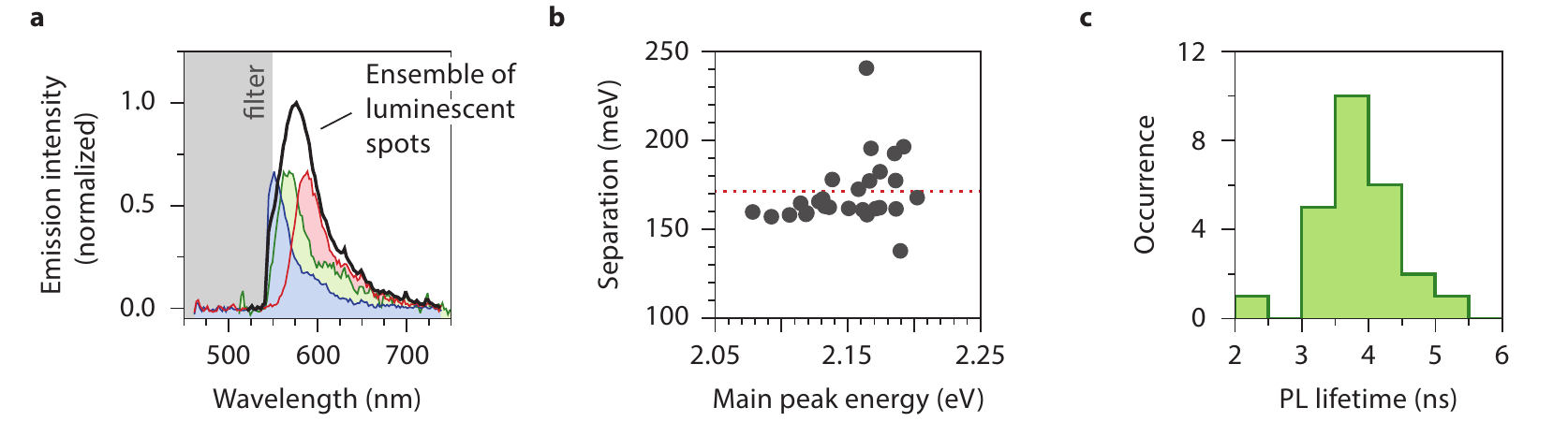}
\caption{
\textbf{The properties of single-photon emitting centres in a borosilicate cover slip.}
\textbf{(a)} Emission spectra of three different individual centres (red, green, blue), as well as the `ensemble' emission spectrum obtained by averaging over 27 centres (black). A long-pass filter at 540 nm cuts the blue side of the spectra. Each individual spectrum has a double-peak shape, but this shape is hidden after ensemble averaging.
\textbf{(b)} There is not a strong correlation between the energy of the main peak, and the separation between the peaks, as obtained from a triple-Gaussian fit (with regular spacing) to spectra of individual centres. The separations are between 150 and 200 meV, while the peak energy varies between 2.07 eV and 2.20 eV.
\textbf{(c)} The fitted PL lifetimes of centres vary between 2 ns and 6 ns, with an average of 3.8 ns.
\label{fig:statistics}
}
\end{figure}

By examining over 20 individual centres in a glass cover slip, we investigated how the PL properties are distributed. The results are presented in Fig.~\ref{fig:statistics}. Fig.~\ref{fig:statistics}a shows the emission spectra of three individual centres (red, green, and blue), as well as the `ensemble emission spectrum' (black) obtained by averaging over measurements on 27 different individual centres. The ensemble spectrum is approximately twice as broad as the individual spectra, indicating a considerable inhomogeneous distribution in the PL spectral properties of the centres. We see that while each individual spectrum seems to consist of several peaks (as in Fig.~\ref{fig:borosilicate}c), this structure is hidden after ensemble averaging. In Fig.~\ref{fig:statistics}b we plot the fitted separation between peaks in the emission spectrum of an individual centre, versus the energy of the main peak. The separations are mostly between 160 and 180 meV, without a correlation with peak energy over this small energy range. The time-resolved PL decay curves of the centres are all nearly single-exponential, as in Fig.~\ref{fig:borosilicate}d. The lifetimes are however distributed between 2 and 6 ns (Fig.~\ref{fig:statistics}c), with an average of 3.8 ns. \\

We confirmed that the luminescent centres observed (Figs.~\ref{fig:borosilicate},\ref{fig:statistics}) are intrinsic to silica, by also investigating a 'fused quartz' cover slip.  The top surface of the quartz slips was difficult to image with the oil immersion setup that we used for borosilicate glass, possibly because the refractive index of borosilicate glass is higher than that of fused quartz. Instead, we used a water immersion objective and a slightly higher laser power (see Methods for details). \\


\begin{figure}[ht]
\centering
\includegraphics[width=\linewidth]{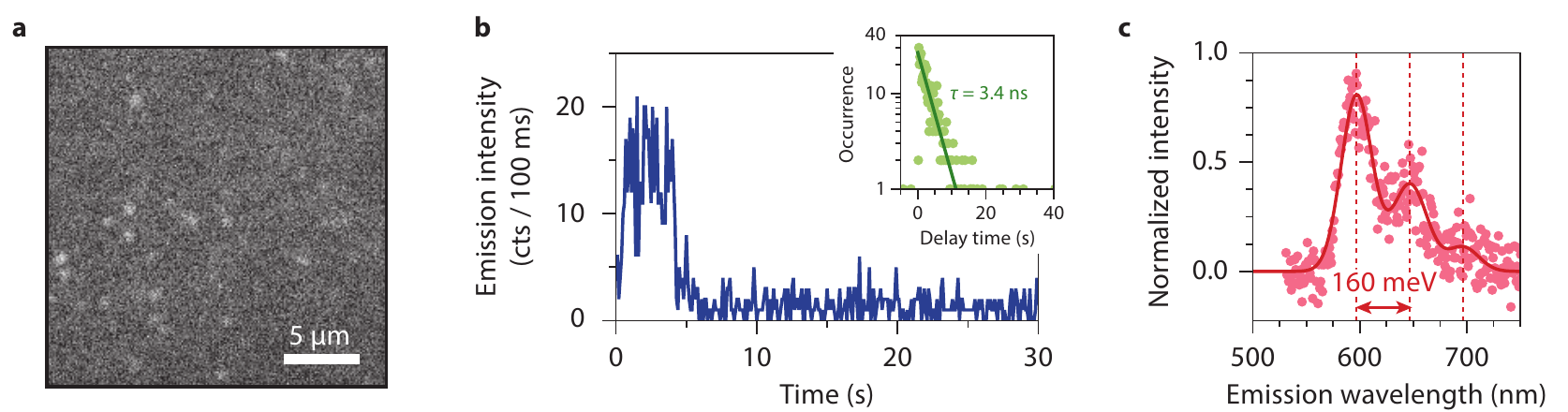}
\caption{
\textbf{Individual defects in quartz cover slips.}
\textbf{(a)} A fluorescence image of a fused quartz cover slip, epi-illuminated with the laser (532 nm; 600 $\mu$W) defocussed to a spot with $\sim$50 $\mu$m diameter.
\textbf{(b)} The intensity trace of an individual spot, that bleaches within 5 s. The PL decay trace during these first 5 s is single-exponential, with a fitted PL lifetime of 3.4 ns.
\textbf{(c)} The emission spectra of the spots in fused quartz have the characteristic multi-peaked appearance. They can be fitted to a series of three Gaussians, with separations of on average 170 meV.
\label{fig:quartz}
}
\end{figure}

The observations on the fused quartz cover slips are summarized in Fig.~\ref{fig:quartz}. There are individual emitting spots with a surface density of 0.01--0.1 $\mu$m$^2$ (Fig.~\ref{fig:quartz}a), although their brightness is considerably lower than that from the defects in borosilicate glass. We conclude that not only borosilicate glass, but also fused quartz contains luminescent defect centres. The centres in quartz bleached typically within a few seconds (Fig.~\ref{fig:quartz}b), probably because we used a high excitation power to get sufficient signal from the weak emitters in the quartz. We were nevertheless able to study the PL characteristics of several individual centres. The PL decay is single exponential, with PL lifetimes of a few nanoseconds (inset of Fig.~\ref{fig:quartz}b). Most importantly, the emission spectrum has the characteristic multiple-peak appearance (Fig.~\ref{fig:quartz}c), although the signal-to-noise is poorer for the quartz defects. The splitting between the emission peaks is very similar to that observed in borosilicate glass, with an average of 170 meV over 17 centres in quartz. The similarities between the PL spectra and lifetime for the two substrates suggests that the centres examined here are intrinsic to silica, appearing both in borosilicate glass and in fused quartz. \\

We further emphasize that the appearance of luminescent defects did not depend on the cleaning procedure of the cover slips. Cover slips taken directly from the box showed characteristics very similar to cover slips that were first cleaned in hydrochloric acid and isopropanol (see Methods). Baking the substrate at 150$^\circ$C for 1 min to remove any organic contaminations adsorbed did not have an effect either. Even surface functionalization with hexamethyldisilazane had no consequences for the density nor the properties of the luminescent defect centres.

\section*{Discussion}

Single-photon emitters with characteristics similar to those we have observed on bare silica substrates (i.e. multi-peaked spectra, ns PL lifetime, non-blinking emission), have been reported before. For example, Wang \emph{et~al.} \cite{Wang2009} reported non-blinking single-photon emission with a PL lifetime of 4--5 nanoseconds and an emission spectrum characterised by multiple peaks separated by $\sim$164 meV. They ascribed it to CdZnSe/ZnSe QDs with a composition gradient at the core--shell interface. These surprising characteristics, inconsistent with the ensemble PL, were explained by a model in which the QDs are permanently positively charged. Our findings here imply that, instead, the PL observed could have originated from luminescent centres in the cover slips used for the experiments in Ref.~\cite{Wang2009}. \\

PL with identical characteristics to those of the centres in silica, has previously also been reported for individual Si QDs \cite{Mason1998,English2002,Martin2008,Kusova2010,Schmidt2012} or SiO$_2$ nanoparticles \cite{Chizhik2009}. While the PL properties of an ensemble of Si QDs depend strongly on the preparation method and on the surface chemistry, individual Si QDs always show PL with the same characteristics (and very similar to those shown in Fig.~\ref{fig:borosilicate},\ref{fig:statistics},\ref{fig:quartz}). In Refs.~\cite{Chizhik2009,Schmidt2012} it was proposed that this PL originates from defects in the amorphous SiO$_2$ surface layer of the QDs, while the much weaker excitonic emission from Si QDs is hard to observe \cite{Schmidt2012}. The peak splitting of (consistently) approximately 160 meV is ascribed to coupling to phonons in SiO$_2$. The same type of defect responsible for emission in Si QDs, may be present at a low density in silica cover slips. Alternatively, the multi-peaked emission spectra ascribed to Si QDs or SiO$_2$ nanoparticles in Refs.~\cite{Mason1998,English2002,Martin2008,Chizhik2009,Kusova2010,Schmidt2012} may in fact have originated from centres in the silica substrates used for the experiments. \\

\begin{figure}[ht]
\centering
\includegraphics[width=\linewidth]{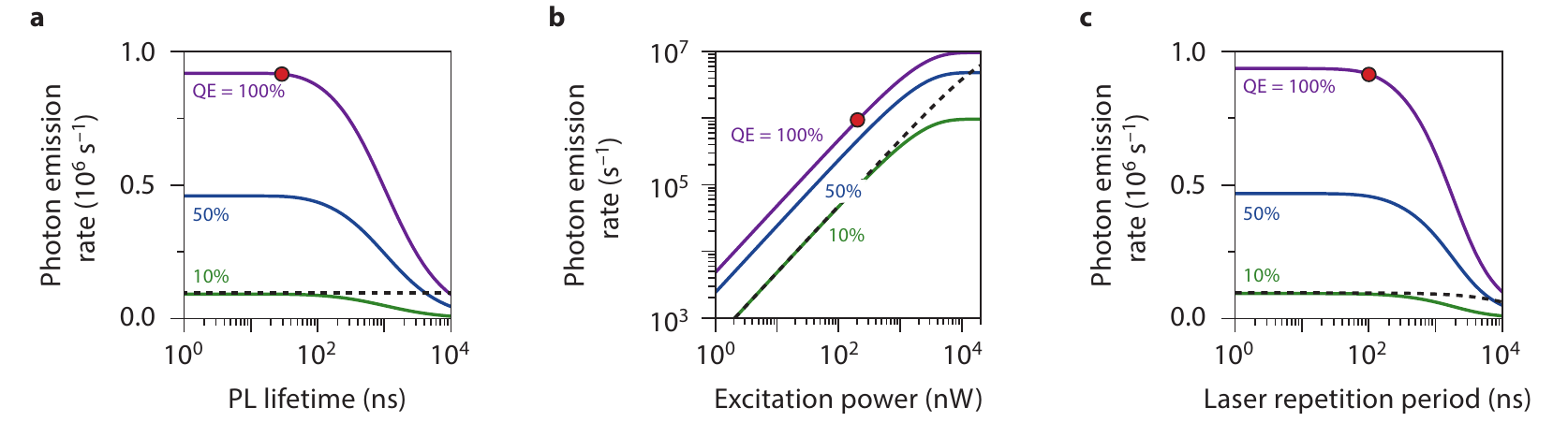}
\caption{
\textbf{Photon emission rates from single-photon emitters.}
We compared the estimated photon emission rates of QDs with an absorption cross-section of $\sigma=10^{-15}$ cm$^2$ and a default PL lifetime of $\tau =$ 30 ns (similar to the dot-in-rods of Ref.~\cite{Rabouw2013}) to centres in silica with $\sigma=10^{-16}$ cm$^2$ and $\tau =$ 5 ns. The default excitation intensity is 200 nW focussed to a spot with a diameter of $532~\textrm{nm}/2$. The default laser repetition period is 100 ns. We consider QDs with quantum efficiencies of 1 (purple), 0.5 (blue), and 0.1 (green). In each panel, the red dot marks the experimental parameters for the dot-in-rods in Ref.~\cite{Rabouw2013}. The dashed line is the emission rate from a luminescent centre in silica. Note that for these calculations we assume a zero quantum efficiency of multi-exciton emission (see Methods). Emitters exhibiting biexciton emission would be brighter under strong excitation than calculated here. One should furthermore realize that in practice under strong excitation, the intensity of emitters may decrease because of charging or bleaching, but these effects are not considered here.
\textbf{(a)} The photon emission rate as a function of the PL lifetime of the QDs.
\textbf{(b)} The photon emission rate as a function of the excitation power.
\textbf{(c)} The photon emission rate as a function of the laser repetition period, at a constant average excitation power of 200 nW. This means that for longer repetition periods, the energy per pulse is higher.
\label{fig:emrates}
}
\end{figure}

In Fig.~\ref{fig:emrates} we compare the estimated brightness of QDs to that of the centres in silica in a single-emitter experiment using a pulsed excitation source of 532 nm, as a function of (a) the PL lifetime of the QDs, (b) the excitation power of the laser, and (c) the repetition period of the laser. See Methods for details of the calculation. We see that with the experimental parameters used in this work (which are the same as in Ref.~\cite{Rabouw2013}, i.e. excitation power = 200 nW, laser repetition period $T=$ 100 ns), a typical QD (red circle) is approximately 10$\times$ as bright as a centre. The emission rate of a centre is estimated at $8\times 10^5$ s$^{-1}$. When considering that the collection and detection efficiency of our setup is approximately 1\%, this value corresponds well to the observation of $6\times 10^3$ counts per second (Fig.~\ref{fig:borosilicate}b). \\

The defect PL from silica is approximately a factor 10 less bright than exciton PL from CdSe/CdS dot-in-rods investigated on the same setup and under the same conditions \cite{Rabouw2013}. In a wide-field image, PL of the centres \emph{in} the silica substrate (Fig.~\ref{fig:borosilicate}a) cannot be distinguished from PL of individual weak emitters intentionally spun-coat \emph{on} the substrate. In our previous work \cite{Rabouw2013}, we were able to properly analyse the CdSe/CdS dot-in-rods (as evidenced by single-emitter characteristics consistent with the ensemble properties), because the dot-in-rods were by far the brightest emitters in the sample. PL from the centres in the substrate became observable (and are even the brightest source of PL) when we attempted to examine CuInS$_2$ QDs. CuInS$_2$ QDs are weaker emitters than CdSe/CdS dot-in-rods, because they have longer PL lifetimes (\textit{viz.} tens to hundreds of nanoseconds) and lower PL quantum efficiencies. In addition, they may have a lower absorption cross-section at 532 nm, and may be more vulnerable to photobleaching. \\

For the experimental parameters used in this work a typical QD is significantly brighter (red circle) than the defects. However, if any of the relevant parameters changes (emitter PL lifetime, quantum efficiency, excitation power, laser repetition rate; see Fig.~\ref{fig:emrates}), the brightness ratio can shift in favour of the luminescent centres. Indeed, since the excited state lifetime of QDs is relatively long-lived and the absorption-cross section is large, the fluorescence signal from QDs saturates more easily than that from the centres in silica. This analysis explains why in experiments on `dim QDs' with a low quantum efficiency, long PL lifetime (such as CuInS$_2$ or Si), and/or poor photostability, the signal from emitting centres in the silica substrate can be the brightest source of PL in the field of observation.

\section*{Conclusion}

Silica cover slips contain intrinsic single-photon emitting centres, at a low surface density of 0.01--0.1 $\mu$m$^{-2}$. The emission spectrum of the centres shows a progression of peaks separated by $\sim$170 meV, with the main peak around 600 nm. The photoluminescence lifetime is a few nanoseconds. The realisation that silica contains such defects is important for the interpretation of spectroscopic experiments on individual emitters when using a silica substrate (glass or quartz). Emission from these defects has probably been observed previously, but was then misinterpreted as originating from the emitters under investigation.

\section*{Methods}

\subsection*{Experimental procedure}

Before measurement, Menzel Gl\"{a}ser 24$\times$24 \# 1 borosilicate cover slips were cleaned by sequential dipping for $\sim$30 s in an HCl bath (36\%), a water bath, and a bath of isopropanol, and then blown dry with N$_2$. Quartz cover slips (Esco Optics, Inc.) were cleaned by bath sonicating in acetone for 15 minutes, soaking in a 1 \% Hellmanex solution at 35$^\circ$C for 30 minutes, soaking in 6 M HCl for 60 minutes, and bath sonicating in methanol for 15 minutes; 5 minutes of bath sonication in Nanopure water occurred between each step. After the final methanol wash, the coverslips were allowed to air dry. \\

Measurements were done with a 10 MHz pulsed laser operating at 532 nm, exciting individual defects with 200 nW focused to a diffraction-limited spot. Excitation and collection of luminescence was done with the same NA = 1.4 100$\times$ oil-immersion objective using Fluka immersion oil `UV transparent fluorescence free' (for borosilicate glass), or with an NA = 1.2 60$\times$ water-immersion objective using deionized water (for fused quartz). Reflected laser light was filtered out with a 540 nm long-pass filter. The fluorescence images were recorded on a Nikon CCD-camera (DS-Qi 1 MC) with an integration time of 300 ms. Time-correlated single photon counting was done with two ID Quantique id100-20 avalanche photo-diodes in a Hanbury-Brown-Twiss setup. Emission spectra are integrated for 5 s (borosilicate glass) or 20 s (fused quartz) on an Acton Research SpectraPro 2300i spectrometer equipped with a PIXIS:100 CCD array by Princeton Instruments.

\subsection*{Estimating the photon emission rate of a single-photon emitter}

We consider a single-photon emitter with absorption cross-section $\sigma$ that is excited with a pulsed laser of average intensity $I$ and repetition period $T$ (where continuous-wave excitation corresponds to the limit $T\rightarrow 0$). The statistics of photon absorption is Poissonian (at least for non-resonant absorption, where the effects of absorption-induced bleach or Stark effects on the energies or strengths of transitions can be neglected). The probability that $n$ photons are absorbed is
\begin{equation}
p(n) = (IT\sigma/\hbar\omega)^n \mathrm{e}^{-IT\sigma/\hbar\omega}/n!,
\end{equation}
where $IT\sigma/\hbar\omega$ is the expectation value for the number of absorption events per pulse.

For our model, we consider emitters with a zero quantum efficiency of multi-exciton emission, as is common for conventional quantum dots. The luminescent centres in silica have zero multi-exciton emission too, as evidenced by the absence of a zero-delay peak in the photon correlation function. One should however keep in mind that high-quality (heterostructured) nanocrystal structures with finite biexciton quantum efficiencies will be somewhat brighter under strong excitation conditions than calculated here. Neglecting the contribution of QD multi-exciton emission has however no effect on the overall conclusion that the defect emission in silica will only be comparable with single-QD PL when the PL lifetime is long or the quantum efficiency of the QD is low.

A zero quantum efficiency of multi-exciton implies that following absorption, the emitter immediately and non-radiatively relaxes to the emissive single-exciton state, irrespective of how many photons are absorbed. The probability that a laser pulse excites the emitter is then
\begin{equation}
\label{eq:exc}
X = 1 - p(0) = 1 - \mathrm{e}^{-IT\sigma/\hbar\omega}.
\end{equation}

After excitation, the probability that the emitter is excited decays exponentially with a characteristic time equal to the excited state lifetime $\tau$. Note that this characteristic time is independent of the number of photons initially absorbed, under the assumption that multi-exciton states have a zero quantum efficiency and therefore immediately relax to the single-exciton state. At time $T$ the next excitation pulse hits, and has again a probability to excite the emitter to its excited state. The excited-state population following the second laser pulse can be higher than after the first laser pulse unless $T$ is much longer than $\tau$. The excited state population evolves by sequential excitation and decay, until a `steady state' situation exists where the decay following laser pulse $i$ is exactly compensated by excitation by laser pulse $i+1$:
\begin{equation}
P \left(1-\mathrm{e}^{-T/\tau}\right) = X \left(1 - P \mathrm{e}^{-T/\tau}\right),
\end{equation} 
where $P$ is the excited state population (i.e. the probability that the emitter is excited) directly following a laser pulse. We can solve that the `steady state' excited-state population directly after a laser pulse is
\begin{equation}
\label{eq:pop}
P = \frac{X}{1+(X-1)\mathrm{e}^{-T/\tau}},
\end{equation}
where $X$ is given by Eq.~\ref{eq:exc}. \\

The photon emission rate averaged over a repetition period $T$ is given by
\begin{equation}
\label{eq:emrate}
\langle\Phi\rangle = \frac{\eta}{\tau} \, \frac{1}{T} \int_0^T P \mathrm{e}^{-t/\tau} \, \mathrm{d}t = \frac{\eta P}{T} \left(1-\mathrm{e}^{-T/\tau}\right),
\end{equation}
where $\eta$ is the quantum efficiency of the emitter, and $P$ is given by Eq.~\ref{eq:pop}. From the general equation for the photon emission rate $\langle\Phi\rangle$ (Eq.~\ref{eq:emrate}) we can obtain approximate expressions in the low-intensity limit ($IT\sigma/\hbar\omega\ll 1$), for continuous wave excitation or for pulsed excitation with a repetition period much longer than the excited state lifetime. Continuous wave excitation corresponds to the limit that $T\rightarrow 0$, i.e. pulses follow each other directly. Then the photon emission rate is
\begin{equation}
\langle\Phi\rangle_\textrm{cw} = \frac{\eta}{\tau} \, \frac{I\sigma/\hbar\omega}{1/\tau+I\sigma/\hbar\omega}.
\end{equation}
The same expression would be obtained from a rate equation model with excitation rate $I\sigma/\hbar\omega$ and decay rate $1/\tau$. Pulsed excitation with a long excitation period corresponds to the limit that $T/\tau \gg 1$:
\begin{equation}
\label{eq:pulsed}
\langle\Phi\rangle_{T\rightarrow\infty} = \eta I\sigma/\hbar\omega,
\end{equation}
i.e. the emission rate no longer depends on the excited state lifetime $\tau$. \\

To estimate the expected brightness of the defects in silica under various experimental conditions, we compare them to the dot-in-rods examined in Ref.~\cite{Rabouw2013}. The brightness of an individual emitter is proportional to the product of PL quantum efficiency $\eta$ and absorption cross-section $\sigma$ (see Eq.~\ref{eq:pulsed}), assuming that the absorption-cross section is sufficiently small ($< 10^{-14}$ cm$^{2}$) that the excitation is not saturated ($\ll 1$ absorption per pulse). For the dot-in-rods investigated in Ref.~\cite{Rabouw2013} the quantum efficiency of the bright state is close to 1 \cite{Lunnemann2013}, and the absorption cross-section at 532 nm is approximately $\sigma = 10^{-15}$ cm$^{2}$ \cite{Leatherdale2002}. The defects in silica are approximately 10 times less bright, so we estimate that for the defects $\eta\sigma\approx 10^{-16}$ cm$^{2}$. For the calculations in Fig.~\ref{fig:emrates} we assume that $\eta=1$ and $\sigma= 10^{-16}$ cm$^{2}$, but the results would change only slightly if the individual values were a little different. We use the full expression for the photon emission rate (Eq.~\ref{eq:emrate}).


\section*{Acknowledgements}

This work is part of the research programme of the `Stichting voor Fundamenteel Onderzoek der Materie (FOM)', which is financially supported by the `Nederlandse Organisatie voor Wetenschappelijk Onderzoek (NWO)'. Financial support from the division of Chemical Sciences (CW) of The Netherlands Organization for Scientific Research (NWO) is acknowledged, under the grant numbers ECHO.712.012.001 (W.v.d.S. and C.d.M.D.) and ECHO.712.014.001 (A.B. and C.d.M.D.) as is the Department of Energy Office of Basic Energy Sciences through Grant DE-FG02-06ER15821 (T.D.K and N.M.B.C). 

\section*{Author contributions statement}

F.T.R. and N.M.B.C. conducted the experiments. A.C.B. and W.v.d.S. provided materials. All authors were involved in the interpretation and discussion of the data. F.T.R. wrote the paper, with input from all other authors.

\section*{Additional information}

\subsection*{Competing financial interests}
The authors declare no competing financial interests.

\end{document}